\documentclass{emulateapj}
\usepackage{graphicx}
\usepackage{natbib}
\usepackage{color}

\begin{document}

\newcommand{\arcm}{$^\prime$}
\newcommand{\arcs}{$^{\prime\prime}$}
\newcommand{\m}{$^{\rm m}\!\!.$}
\newcommand{\D}{$^{\rm d}\!\!.$}
\newcommand{\F}{$^{\rm P}\!\!.$}
\newcommand{\kms}{km~s$^{-1}$}
\newcommand{\ks}{km~s$^{-1}$}
\newcommand{\ms}{M$_{\odot}$}
\newcommand{\rs}{R$_{\odot}$}

\newcommand{\hip}{$Hipparcos$}
\newcommand{\ond}{Ond\v{r}ejov}
\newcommand{\ova}{Ostrava}
\newcommand{\valmez}{Vala\v{s}sk\'e Mezi\v{r}\'{\i}\v{c}\'{\i}}
\newcommand{\jil}{J\'{\i}lov\'e}
\newcommand{\hra}{Hradec Kr\'alov\'e}
\newcommand{\pec}{Pec pod Sn\v{e}\v{z}kou}
\newcommand{\La}{La~Silla}

\title{The first analysis of extragalactic binary-orbit precession.\thanks{Based on data
collected with the Danish 1.54 m telescope at the ESO La Silla Observatory. 
  } }

\author{P. Zasche~
    \and M. Wolf~         }
    
\affil{Astronomical Institute, Charles University in Prague, Faculty of Mathematics and Physics, CZ-180~00 Praha 8, \\
             V~Hole\v{s}ovi\v{c}k\'ach 2, Czech Republic, email: zasche@sirrah.troja.mff.cuni.cz}

\date{Received \today}


\begin{abstract}
 {The main aim of the present paper is the very first analysis of the
binary-orbit precession out of our Galaxy. }
 {The light curves of an eclipsing binary MACHO 82.8043.171 in the Large Magellanic cloud (LMC)
were studied in order to analyse the long-term evolution of its orbit.}
 {It is a detached system that is undergoing rapid orbit precession. The inclination of the orbit towards
the observer has been changing, which has caused the eclipse depth to become lower over the past
decade, and this is ongoing. The period of this effect was derived as only about 77 years, so it is
the second fastest nodal motion known amongst such systems nowadays. This is the first analysis of
an extragalactic binary with nodal precession. This effect is probably caused by a distant third
body orbiting the pair, which could potentially be detected via spectroscopy.}
 {Some preliminary estimates of this body are presented. However, even such a result can tell us
something about the multiplicity fraction in other galaxies.}
 \end{abstract}



\section{Introduction}

After more than a century of intensive study of eclipsing binaries (hereafter EBs), they still
represent the best method for deriving the masses, radii, and luminosities of stars. Thanks to
modern (ground- and space- based) telescopes, we are able to also study these objects in other
galaxies and to apply the same methods as used in our solar neighbourhood. Nevertheless, there is
still a difference in the precision of EB parameters as derived for Galactic ($\sim$2-3\%) and
extragalactic ($\sim$10\%) eclipsing binaries (see e.g. \cite{2004NewAR..48..679C},
\cite{2004NewAR..48..731R}).

The extragalactic EBs can serve as an independent tool for deriving galactic properties and
also help us answer such important questions as "Is the chemical composition of our Galaxy the same as
the neighbouring ones?" or "What is the binary and multiplicity fraction in other galaxies?"
Studying other galaxies via detailed analysis of individual stars can provide some useful hints
for answering these questions.

EBs are quite common, even in some nearby galaxies \citep{2006A&A...459..321V}. However, one
special group of EBs is still rather rare - those undergoing an orbit precession. If the
orientation of the EB orbit is moving in space, then the depths of eclipse also change, and we can
detect this orbit precession. Observing the binary at different time epochs can help us to derive
the inclination towards the observer as a function of time. This effect is usually caused by the
third component orbiting the close pair \citep{1975A&A....42..229S}. However, we still know
of only a few such systems, and detailed analysis has only been carried out for those
located in our Galaxy at present. This is the first time such an effect has been studied in an
extragalactic source.

\section{The system MACHO~82.8043.171}

The Large Magellanic Cloud (LMC) is a close galaxy, which has been observed quite frequently during
the past decades. There have been two major photometric surveys of LMC stars, MACHO
\citep{2007AJ....134.1963F} and OGLE \citep{2011AcA....61..103G}, while discovering a huge number
of variable stars including the EBs. The MACHO survey lasted from 1993 to 1998 and the OGLE from
2002 to 2008. Quite surprisingly, thanks to these two surveys we know more eclipsing binaries in
the LMC than in our Galaxy \citep{2011AcA....61..103G}. However, owing to the low declination of
LMC stars, there are still many interesting systems that lack detailed analysis.

The object called MACHO~82.8043.171 (= OGLE-LMC-ECL-17359, $V=16.98$~mag) was observed by both
photometric surveys, so we can harvest the databases for a complete light curve analysis.  The
system is a detached eclipsing binary with its short orbital period of about 1.26 days
\citep{2011AcA....61..103G}. According to its photometric indices (see below), it is probably a
B2V-type system. Our new observations were obtained during a four-month period in the 2012/2013
season, using the 1.54-meter Danish telescope located at the La Silla observatory in Chile
(hereafter DK154), but operated remotely from the Czech Republic. The standard Cousins filter $I$
was used for our new observations, in agreement with the OGLE survey. Therefore, we can make the
first analysis of this interesting system ranging over two decades to detect some long-term
changes.

For a complete light curve analysis, we need up-to-date ephemerides of the binary. Using the MACHO
and OGLE photometry and deriving the precise times of minima, the following ephemerides were used
for the light curve analysis: $$ \mathrm{Prim.Min.} = \mathrm{HJD}\,\,\, 24\,\,53901.4155 +
1.\!\!^\mathrm{d}25652350 \cdot E.$$ These ephemerides are also suitable for planning future
observations. From the observations we determine that the orbit is circular (i.e. no deviation of
secondary minima appears).

\begin{figure}
  \centering
  \includegraphics[width=85mm]{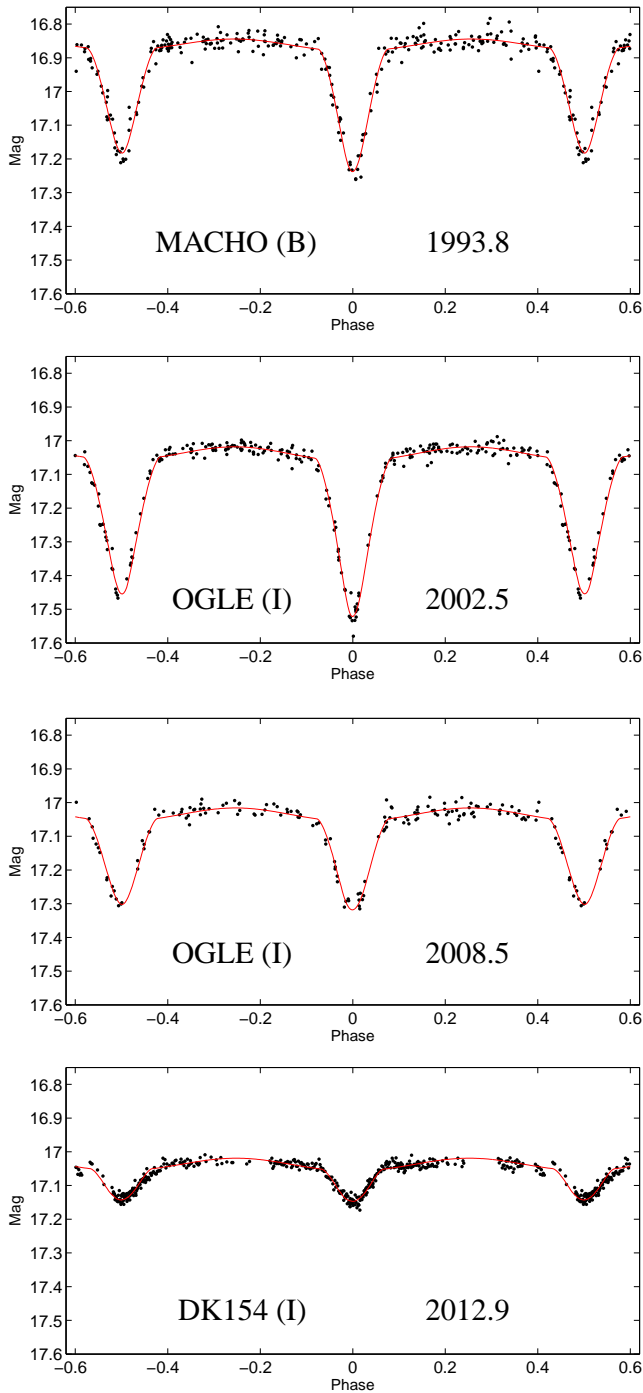}
  \caption{Sample of plots of the light curves at different epochs. Since the range for
  the y-axis is the same for all plots, the change in amplitude is clearly visible.}
  \label{Fig1}
\end{figure}

\section{Analysis}

The light curve fitting of available photometric data was performed using the program PHOEBE, ver.
0.31a \citep{2005ApJ...628..426P}, which is based on the Wilson-Devinney algorithm
\citep{1971ApJ...166..605W} and its later modifications.

The following procedure was used for the analysis. First, the ephemerides and temperature of the
primary component were fixed for the entire computational process. The temperature of the primary
component was estimated from its photometric index. Owing to many different sources and a rather
wide range of magnitude values for the red and infrared filters, this approach was found to be
problematic. For example, the $(V-R)$ photometric index ranges from $-$0.60~mag
\citep{2004AAS...205.4815Z} 
to 0.065~mag \citep{2007ApJ...663..249D}, which yields a range of spectral types from O to A. On
the other hand, \cite{2000A&A...364..455L} published the Str\"{o}mgren $uvby$ photometry, which can
be transformed \citep{2001A&A...369.1140H} into the Johnson $UBV$ system. After these
transformations, the unreddened values of the photometric indices $(B-V)_0 = -0.23$~mag resulted,
as did $(U-B)_0 = -0.87$~mag, which clearly show the star to be about a B2V spectral type
\citep{1974ASSL...41.....G}. Although the star is not single, the two components are rather similar
(see below), so we accepted this estimation. The resulting value of $E(B-V) = 0.42$~mag was quite
surprising, because it is a bit larger than commonly used for LMC binaries, but it is still
acceptable \citep{2000A&A...364..455L}. Another source of Johnson magnitudes is, for example, the
one by \cite{2004AJ....128.1606Z}, who published the $UBVI$ photometry. Regrettably, this
photometry is also unusable due to larger errors and unacceptable $(U-B)_0$ values. To conclude,
after assuming the B2V spectral type, we fixed the temperature at $T_1 = 21000~$K
\citep{2011ApJS..193....1W} for the computing process.

Owing to rather different quality of the individual light curves, the first OGLE data set (2002.5)
was used as the initial one. With this light curve we analysed the system, resulting in a set of
parameters for both components, see Table \ref{TabLC}. These parameters are the best we were able
to derive from the available light curves. The values for temperature, the secondary component,
Kopal's modified potential, luminosities, etc. were used for the subsequent light curve analysis of
data obtained at different epochs. However, the lack of other relevant information (e.g. from
spectroscopy) meant that some of the parameters have to be fixed for the whole analysis. We assumed
a circular orbit (i.e. $e = 0$) and a mass ratio $q = 1$. The albedo coefficient remained fixed at
value 1.0, gravity darkening coefficients at $g = 1.0$, and the synchronicity parameters  at$F =
1$. The limb darkening coefficients were interpolated from the van~Hamme's tables
\citep{vanHamme1993}. No third light was detected for any of the light curves.

This analysis is based on the assumption that the two components are rather similar to each other.
This presumption was derived from the resulting parameters from Table \ref{TabLC} (similar
temperatures and luminosities), as well as from the $(B-R)$ photometric index at various orbital
phases as derived from the MACHO data. On the other hand, the use of photometry by
\cite{2000A&A...364..455L} was obtained during many nights of observations, which did not take the
current orbital phase of the binary into account, which means that some of the photometric data
points could have been obtained during the eclipses. However, no other better photometry is
available, and owing to the duration of eclipses (both about 1/10 of the orbital period), only
about 20\% of the data points are likely to be influenced by the eclipses. However, we still
believe that this does not play a significant role because of the similarity of the two eclipsing
components. The best way would be to obtain the individual times of observations for the photometry
of \cite{2000A&A...364..455L}, but after communicating with the author, this information is no
longer available.

During the fitting process, the mass ratio can also be fitted. As a result, we made this attempt,
but it did not result in any significant improvement of the fit. The mass ratio is only poorly
constrained here, which agrees with a previous finding that detached eclipsing binaries with only
partial eclipses are not suitable for deriving the mass ratio only from the light curves, see e.g.
\cite{Terrell2005}. We can therefore only roughly estimate the uncertainty of the mass ratio to be
about 0.1.

\begin{figure}
  \centering
  \includegraphics[width=85mm]{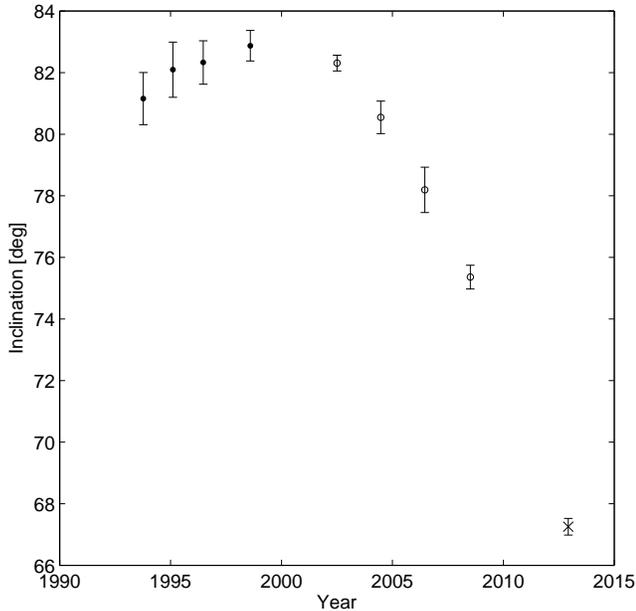}
  \caption{Plot of the changing inclination as a function of time. For the explanation of the symbols
  see Table \ref{TabIncl}.}
  \label{Fig2}
\end{figure}

A sample of fitted light curve plots at different time epochs is given in Fig. \ref{Fig1}. As one
can see, the depths of both primary and secondary minima are changing over the two decades. From
these fits, the individual inclination angles as derived from the Wilson-Devinney algorithm are
given in Table \ref{TabIncl} and plotted in Fig.\ref{Fig2}. The inclination is seent to
change quite fast (more than 2$^\circ$ every year), and the amplitude of photometric variation is
rather shallow at present.

One can also ask how we dealt with the different light curves in different filters for the complete
analysis. Our new observations were obtained in the same filter ($I$) as the OGLE survey. The
second OGLE filter ($V$) was not used because of a limited dataset. On the other hand, both the $B$
and $R$ filters from the MACHO survey were used for the analysis. The light curves in different
filters and different epochs were analysed separately, resulting in different inclination angles.
These two values of inclination angles (but obtained during the same epoch) were averaged into the
value presented in Table \ref{TabIncl}. We could afford to combine different filters and
instruments for the analysis, because the different luminosity levels for different passbands were
also computed.

\begin{table}
 \centering
 \begin{minipage}{84mm}
  \centering
  \caption{The parameters of the light curve.}
  \label{TabLC}
  \begin{tabular}{@{}c c c @{}}
\hline
   Parameter  & Value & Error \\
 \hline
  $T_1$ [K]  & \multicolumn{2}{c}{21000 (fixed)} \\
  $T_2$ [K]  & 20040 & 260   \\
  $\Omega_1$ & 5.426 & 0.048 \\
  $\Omega_2$ & 4.548 & 0.031 \\
  $L_1$ [\%] & 41.0 & 2.4    \\
  $L_2$ [\%] & 59.0 & 2.6    \\
  $r_1/a$    & 0.267 & 0.005 \\
  $r_2/a$    & 0.284 & 0.006 \\
 $q=M_2/M_1$ & \multicolumn{2}{c}{1.00 (fixed)} \\
 $e$         & \multicolumn{2}{c}{0.00 (fixed)} \\
 $F_1=F_2$   & \multicolumn{2}{c}{1.00 (fixed)} \\
 $A_1=A_2$   & \multicolumn{2}{c}{1.00 (fixed)} \\
 $g_1=g_2$   & \multicolumn{2}{c}{1.00 (fixed)} \\
 \hline
\end{tabular}
\end{minipage}
\end{table}

\begin{table}
 \centering
 \begin{minipage}{84mm}
  \centering
  \caption{The inclination angles as derived from different light curves. Mean epochs for each data set
  are given.}  \label{TabIncl}
  \begin{tabular}{@{}c c c c c@{}}
\hline
  &  & \multicolumn{2}{c}{Inclination [deg]} & Fig.\ref{Fig2} \\
 Year   & Source & Value & Error & Symbol \\
 \hline
 1993.770  &  MACHO  & 81.16 &  0.85  & \Huge{$\cdot$} \\[-1.4mm]
 1995.104  &  MACHO  & 82.09 &  0.90  & \Huge{$\cdot$} \\[-1.4mm]
 1996.473  &  MACHO  & 82.33 &  0.70  & \Huge{$\cdot$} \\[-1.4mm]
 1998.594  &  MACHO  & 82.87 &  0.50  & \Huge{$\cdot$} \\
 2002.504  &  OGLE   & 82.31 &  0.26  & $\circ$  \\
 2004.474  &  OGLE   & 80.55 &  0.53  & $\circ$  \\
 2006.459  &  OGLE   & 78.19 &  0.74  & $\circ$  \\
 2008.510  &  OGLE   & 75.36 &  0.38  & $\circ$  \\
 2012.925  &  DK154  & 67.25 &  0.27  & $\times$ \\
 \hline
\end{tabular}
\end{minipage}
\end{table}

\section{Results}

The system undergoes a nodal precession of its orbit, which is probably caused by an orbiting third
body. This effect of binary orbit precession is nothing new; however, it has been observed and analysed
for the first time for an extragalactic source. One can compute the nodal period from the equation
given in \cite{1975A&A....42..229S}:
$$P_{\mathrm{nodal}} = \frac{4}{3} \left( 1+ \frac{M_1 + M_2}{M_3} \right) \frac{P_3^2}{P} (1-e_3^2)^{3/2}
\left( \frac{C}{G_2} \cos j \right)^{-1},$$ where subscripts 1 and 2 stand for the components of
the eclipsing binary, while 3 stands for the third body. The term $C$ is the total angular momentum
of the system, while ${G_2}$ stands for the angular momentum of the wide orbit.

Unfortunately, we are not able to derive the nodal period using this equation owing to unknown
individual orbital parameters and masses of the components. We therefore used a simplified approach to
fitting the term ``$\cos i$", as given, say, in \cite{1994A&A...284..853D}:
$$\cos i = \cos I \cdot \cos i_1 - \sin I \cdot \sin i_1 \cdot \cos (2\pi(t-t_0)/P_{\mathrm{nodal}}), $$ where $I$
is the inclination of the invariant plane against the observer's celestial plane, $i$ is the
inclination of the eclipsing binary, and $i_1$ is the inclination between the invariant plane and
the orbital plane of the eclipsing binary.

\begin{figure}
  \centering
  \includegraphics[width=85mm]{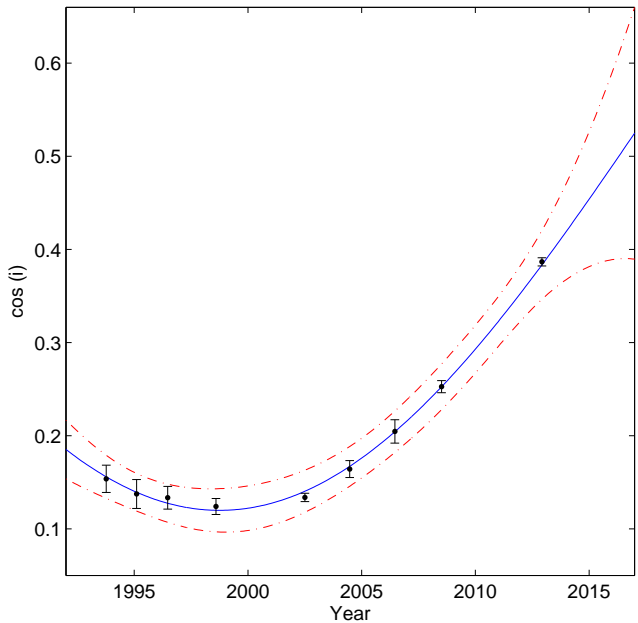}
  \caption{Plot of the "$\cos i$" term and its final fit (see the text). A confidence level of 95\% is shown with the dash-dotted line.}
  \label{Fig3}
\end{figure}

 Figure \ref{Fig3} shows the result of our fitting. The resulting nodal period is only about 76.9 $\pm$
10.1 years. However, because of the poor coverage of this period with only two decades of data, this
result is still rather preliminary. New and more precise observations (both photometry and
spectroscopy) are needed in upcoming years. The resulting period of nodal precession is the second
shortest among known systems to date, the fastest motion being that of the well-known system
V907~Sco \citep{1999AJ....117..541L} with its nodal period about 68~years. For MACHO~82.8043.171, we
do expect that the photometric eclipses will stop as late as about 2017. Until that time only very
shallow ellipsoidal variations of the order of 0.03 mag (I filter) remain.

As a by-product we also derived the inclination angles $I$ and $i_1$ from the equation for $\cos
i$. These two quantities resulted in $I = 41.1^\circ \pm 11.8^\circ$ and $i_1 = 42.0^\circ \pm
13.2^\circ$. The values define the orientation of the system in space and towards the observer (see
Fig. 2 in \citealt{1975A&A....42..229S}). Both these angles could potentially be used for future
dynamical studies, should the third-body orbit be discovered via spectroscopy.

\section{Discussion}

Discovering the nodal precession of MACHO~82.8043.171, one can ask whether the third body causing
this effect is detectable with current data or facilities. The easiest method is the light curve
analysis and detection of the third light. However, no such additional light was discovered, so that
it gives some constraints on this body. Assuming a detection limit of about 1\% of the total light,
then the undiscovered third component has to be spectral type A2 or later, assuming it lies on
the main sequence. We can only speculate about its period and semimajor axis, so that the
amplitudes of radial velocity variations are also questionable. The absence of any third light makes it
similar to the recently discovered system HS~Hya \citep{2012A&A...542L..23Z}, where the third body
causing the nodal precession of the eclipsing pair also cannot be detected in the light curve
solution, but was discovered via spectroscopy.

\begin{figure}
  \centering
  \includegraphics[width=85mm]{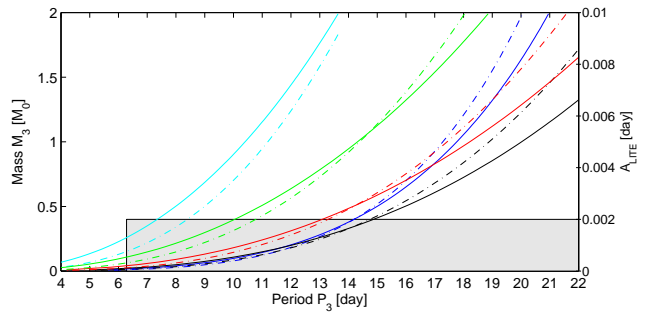}
  \caption{Plot of the predicted parameters of the third body resulting from the nodal period. Its
  period (the x-axis) versus its mass (y-axis, solid curves) is computed from the nodal period, and
  the corresponding amplitude of the light-time effect was computed (y-axis on the right, dash-dotted
  curves). The shadowed area represents the possible parameters. Different colours represent different
  orientations of the orbit in space. See the text for details.}
  \label{Fig4}
\end{figure}

Moreover, the configuration of the system has to be hierarchical because of its stability
\citep{1992ASPC...32..212H}. Another limiting factor is the fact that any variation in the times of
minima is detectable in the $O-C$ diagram of MACHO~82.8043.171 (also analogous to HS~Hya). As a
result, a detection limit of about 0.002~days also yields some constraints on the third-body orbit,
mainly the period, see e.g. \cite{Mayer1990}. Using the equation introduced in
\cite{1975A&A....42..229S}, and applying many ad hoc assumptions (e.g. orientation of the orbit in
space, fixing the eccentricity to zero), we can roughly estimate, for example, the orbital period
of the third component. See Fig.\ref{Fig4} for some results, where the predicted masses (solid
curves) and amplitudes of the light-time effect (dash-dotted curves) are plotted with respect to
the orbital period $P_3$. The individual colours stand for different inclinations of the orbits:
90$^\circ$ (blue), 70$^\circ$ (black), 50$^\circ$ (red), 30$^\circ$ (cyan), and 10$^\circ$ (green),
respectively. Moreover, exists the dynamical stability criterion also exists, and it gives the
lower limit of the third-body period: $P/P_3 > 5$, see e.g. \cite{2008MNRAS.389..925T}, resulting
in a minimum period of about 6.28~days. Considering all these criteria, the shadowed area in
Fig.\ref{Fig4} shows the most probable solutions. The expected period $P_3$ should probably be from
6 to 15 days. Moreover, as we can see from Fig. \ref{Fig4}, the limit of no detectable period
variation of the order of 0.002~days gives a better constraint on the mass of the third body (about
0.4~M$_\odot$) than the absence of the third light, which yields an upper limit of mass of about
2~M$_\odot$. Such a companion is therefore probably an M-dwarf star.

Besides MACHO~82.8043.171, we currently know of only four other systems with derived nodal periods.
However, this unique system is the first analysed eclipsing binary with changing inclination
outside our own Galaxy. The authors are aware of the most important deficiency of the present
analysis, which is the lack of radial velocity measurements, or the detailed spectroscopic study
discovering the third component. On the other hand, as we can see, for example, in the system of HS~Hya, the
third body could have a period of hundreds of days, so to discover it one needs spectroscopic
monitoring over several months. Moreover, other EBs within the LMC are much brighter and also have
longer periods. There is still no radial velocity study of an LMC eclipsing binary with such a
short orbital period. Precise spectroscopic observations for such a faint target would only be
possible using 4 m class telescopes or even larger. This first analysis of MACHO~82.8043.171 could
serve as a starting point for other astronomers to initiate observing campaigns or to submit
observing proposals for this target on large telescopes.

\section{Conclusion}

More detailed study of such systems would potentially be very important for several reasons. First,
EBs are still the best method for deriving precise masses and radii of stars, and also for
calibrating the cosmic distance ladder. Secondly, the chemical compositions of such systems should
be studied in order to compare the LMC and our Galaxy. There are some traces of different
composition between LMC, SMC, and our Galaxy stars, which were derived using EBs
\citep{2004NewAR..48..731R}. Additionally, the EBs serve as independent distance indicators to the
LMC (up to 2\%, \citealt{Pietr2013}). And finally, the changing inclination indicates that there is
a hidden component orbiting this EB pair, which could tell us something about the stellar
multiplicity of the LMC in general. Observing the suspicious EBs would help us discover these third
bodies, which is otherwise rather complicated for such distant objects. (Spectroscopy is
time-consuming and magnitude-limited, and interferometry cannot be used for Magellanic clouds.)

Additionally, multiple systems with moving orbital planes are ideal astrophysical laboratories for
dynamical studies. The observable quantities can be directly compared with theoretical models.
Therefore, each new system is very promising. In our Galaxy we know of 11 such systems nowadays
\citep{2012A&A...542L..23Z}. \cite{2011AcA....61..103G} noted 17 systems in the LMC, and from the
KEPLER data there were seven more of these binaries \citep{2013ApJ...768...33R}.


\medskip
\begin{acknowledgements}

An anonymous referee is acknowledged for the useful comments and suggestions that
significantly improved the paper. Dr. William Hartkopf is also gratefully acknowledged for his help
with the language level of the manuscript. We thank the MACHO and OGLE teams for making all of
the observations publicly available. We are also grateful to the ESO team in La Silla for
their help in maintaining and operating the Danish telescope. This work was supported by the Czech
Science Foundation grant no. P209/10/0715, by the grant UNCE 12 of Charles University in
Prague, and by the grant LG12001 of the Ministry of Education of the Czech Republic. This research made use of the SIMBAD database, operated at the CDS, Strasbourg, France, and of NASA's
Astrophysics Data System Bibliographic Services.

\end{acknowledgements}

\end{document}